\documentclass[aps,prd,floats,floatfix,twocolumn,nofootinbib]{revtex4}

\usepackage{graphicx}
\usepackage{bm}

\newcommand{\third}{{\scriptstyle\frac{1}{3}}}
\newcommand{\twothirds}{{\scriptstyle\frac{2}{3}}}
\newcommand{\half}{{\scriptstyle\frac{1}{2}}}
\newcommand{\fourth}{{\scriptstyle\frac{1}{4}}}

\newcommand{\eqref}[1]{(\ref{#1})}
\newcommand{\beq}{\begin{equation}}
\newcommand{\eeq}{\end{equation}}
\newcommand{\beqa}{\begin{eqnarray}}
\newcommand{\eeqa}{\end{eqnarray}}
\newcommand{\hs}{\hspace{0.075em}}

\begin{document}

\title{Controlling the Growth of Constraints in Hyperbolic
Evolution Systems}

\author{Lee Lindblom${}^1$, Mark A. Scheel${}^1$,
Lawrence E. Kidder${}^2$,\\ Harald P. Pfeiffer${}^1$, 
Deirdre Shoemaker${}^2$, 
and Saul A. Teukolsky${}^2$}
\affiliation{${}^1$ Theoretical Astrophysics 130-33, California Institute of
Technology, Pasadena, CA 91125}
\affiliation{${}^2$
  Center for Radiophysics and Space Research, Cornell University,
  Ithaca, New York, 14853}

\date{\today}

\begin{abstract}
Motivated by the need to control the exponential growth of constraint
violations in numerical solutions of the Einstein evolution equations,
two methods are studied here for controlling this growth in general
hyperbolic evolution systems.  The first method adjusts the evolution
equations dynamically, by adding multiples of the constraints, in a
way designed to minimize this growth.  The second method imposes
special constraint preserving boundary conditions on the incoming
components of the dynamical fields.  The efficacy of these methods is
tested by using them to control the growth of constraints in fully
dynamical 3D numerical solutions of a particular representation of the
Maxwell equations that is subject to constraint violations.  The
constraint preserving boundary conditions are found to be much more
effective than active constraint control in the case of this Maxwell
system.
\end{abstract}


\maketitle

\section{Introduction}
\label{s:introduction}

Perhaps the most critical problem facing the numerical relativity
community today is the exponential growth of constraints in evolutions
of black hole spacetimes.  The evolution equations guarantee that
constraints that are satisfied exactly on a spacelike surface will be
satisfied throughout the domain of dependence of that surface.
However, this result does not guarantee that small initial violations
of the constraints will remain small, or that constraint violations
will not be injected into the computational domain through timelike
boundaries.  Experience has shown that constraint violations tend to
grow exponentially in the numerical evolution of black hole spacetimes
({\it e.g.},~\cite{Kidder2001,Lindblom2002,Scheel2002}). These {\it
constraint violating instabilities} have been shown to be numerically
convergent and thus represent unstable solutions to the partial
differential equations.  At present these instabilities are the
limiting factor that prevents these numerical simulations from running
for the needed time with the required accuracy.

A variety of approaches have been explored in a number of attempts to
control the growth of the constraints:

1) Fully constrained evolution, in which the constraint equations are
re-solved periodically ({\it e.g.} at each time step) have been used with
great success in spherically symmetric and axisymmetric
problems~\cite{StPi85,AbEv92,ch93,AbEv93,ACST94,Choptuik2003,
  Choptuik2003a}.  These methods have not gained widespread use in 3D
simulations, however, due in part to the high cost of solving the
elliptic constraint equations.  Difficult questions also remain
unresolved for this method about the proper boundary conditions to
impose on the constraint equations at black hole excision boundaries.
With the development of more efficient elliptic solvers and the
absence of a better alternative however, fully constrained methods are
starting to be developed and tested in 3D now as
well~\cite{Anderson2003,BGGN03,Schnetter2003}.

2) Auxiliary dynamical fields have been introduced into the system
whose evolution equations are designed to drive the constraints toward
zero~\cite{Brodbeck1999}.  This technique has the disadvantage that it
requires the size of the dynamical system to be significantly
expanded.  It has not been tested extensively, but the first numerical
results were not uniformly successful~\cite{Siebel2001,Yoneda2001}.

3) More sophisticated boundary conditions have been introduced whose
purpose is to control the influx of constraint violation through the
timelike boundaries of the computational domain~\cite{Stewart1998,
  FriedrichNagy1999, Calabrese2002a, Szilagyi2002, Calabrese2003,
  Szilagyi2003, Frittelli2003a, Frittelli2003b, Calabrese2003a,
  Frittelli2003c}. This approach seems very promising, although the
current methods may not be fully compatible with the physical
requirement that waves pass through the boundaries without reflection.
Further these boundary condition methods may not completely solve the
constraint violating instability problem in systems like the Einstein
evolution equations, where constraint violations are driven both by
bulk and by boundary terms in the equations.  But this technique can
(as we will demonstrate below) significantly improve the influx of
constraint violations through the timelike boundaries of the
computational domain.

4) Dynamically changing the evolution equations, through the addition
of terms proportional to the constraints, has been proposed as a way
to minimize constraint growth.  This method (developed by Manuel
Tiglio and his collaborators~\cite{Tiglio2003a,Tiglio2003b}) has had
some success in controlling the growth of constraints in simple
numerical solutions of the Einstein evolution equations.  We find that
this technique when used in combination with standard boundary
conditions is not effective however in controlling the influx of
constraint violations through the boundaries of the computational
domain in fully dynamical situations.

In this paper we explore in some detail two of these methods for
controlling the growth of constraint violations in hyperbolic
evolution systems.  First, we develop a refined version of the
dynamical constraint control method being used by Tiglio and
collaborators~\cite{Tiglio2003a,Tiglio2003b}. In particular we
introduce a more natural norm on the constraints, which has the
property that its evolution can be predicted numerically with greater
accuracy.  We expect that dynamical constraint control based on this
new constraint norm should be more stable and robust than the current
method.  And second, we explore the use of constraint preserving
boundary conditions.  In this method (explored previously by
Calabrese and collaborators~\cite{Calabrese2002a,Calabrese2003a}) the
constraints are decomposed into characteristic ingoing and outgoing
fields of the constraint evolution equations.  Setting the incoming
components of the constraint fields to zero provides boundary
conditions for the incoming parts of the dynamical fields of the
principal evolution system.  We test both of these methods by applying
them to a non-trivial hyperbolic evolution system (a particular
representation of the Maxwell system~\cite{Kidder2002,Reula2002}) that
is analogous to, but much simpler than, the Einstein evolution system.
Our tests---using fully dynamical time dependent solutions on domains
with open boundaries---reveal that the constraint preserving boundary
conditions are much more effective than active constraint control for
this Maxwell system.  Some features of this system are rather special,
and it is possible that in more generic systems (like the Einstein
equations) the active constraint control method may be complimentary
to the constraint preserving boundary condition method.

We define and review in Sec.~\ref{s:maxwell} the particular form of
the Maxwell evolution system~\cite{Kidder2002,Reula2002} that we use to
illustrate and test the constraint control methods studied here.  We
refer to this system as the ``fat'' Maxwell system since it replaces the
usual representation of the Maxwell system, which has six independent
field components, with a representation having twelve.  We also present
in Sec.~\ref{s:maxwell} the decomposition of the dynamical fields used
in this fat Maxwell system into characteristic parts.  In
Sec.~\ref{s:constraintcontrol} we develop the equations needed to
perform active constraint control, in particular on the fat Maxwell
evolution system.  We determine the constraint evolution equations for
this system, and derive an improved norm on the constraint fields.  We
show how the evolution of this new constraint norm should generically
be more accurately determined (and hence should provide better
constraint control) in numerical solutions.  In
Sec.~\ref{s:boundaryconditions} we develop the particular form of
constraint preserving boundary conditions studied here. We present the
decomposition of the constraint fields into characteristic parts, and
show how these can be used to provide boundary conditions for the
principal evolution system.  Finally in Sec.~\ref{s:numericalresults}
we use these methods to control the growth of constraints in fully
dynamical 3D numerical evolutions of the fat Maxwell evolution system.
We note that both the active constraint control mechanism and the
constraint preserving boundary conditions developed here are
applicable to rather general hyperbolic evolution systems.  We focus
our discussion on the fat Maxwell system in order to make the analysis
less abstract, and to provide a simple system on which to perform
numerical experiments.

\section{Fat Maxwell Evolution System}
\label{s:maxwell}

Our primary interest here is to understand how to control the growth
of constraints in hyperbolic evolution systems.  We will focus our
attention on quasi-linear systems of the form,
\begin{equation}
\partial_t u^\alpha + A^{k\,\alpha}{}_\beta\partial_k u^\beta = F^\alpha,
\label{e:evolsystem}
\end{equation}
where $u^\alpha$ represents the dynamical fields, and
$A^{k\,\alpha}{}_\beta$ and $F^\alpha$ may depend on $u^\alpha$ but
not its derivatives.  We assume that the evolution system described in
Eq.~\eqref{e:evolsystem} is also subject to a set of constraints,
$c^A=0$, which we assume have the general form
\begin{equation}
c^A=K^{A\,k}{}_\alpha\partial_k u^\alpha + L^A,\label{e:constraintdefs}
\end{equation}
where $K^{A\,k}{}_\beta$ and $L^A$ may depend on the dynamical fields
$u^\alpha$ but not their derivatives.  We assume that these constraints
are preserved as a consequence of the evolution equations.  In particular
we assume that the constraints satisfy an evolution equation of the form
\begin{equation}
\partial_t c^B + A^{k\,B}{}_D\partial_k c^D = F^B{}_Dc^D,
\label{e:constsystem}
\end{equation}
where $A^{k\,B}{}_D$ may depend on the dynamical fields $u^\alpha$,
while $F^B{}_D$ may depend on $u^\alpha$ and its spatial derivatives
$\partial_ku^\alpha$.  When this constraint evolution system is
hyperbolic the constraints will remain satisfied within the domain
of dependence of the initial surface if they are satisfied initially.
We note that multiples of the constraints of the form given in
Eq.~\eqref{e:constraintdefs} may be added to the principal evolution
system Eq.~\eqref{e:evolsystem} without changing the physical
(constraint satisfying) solutions of the system
or the basic structure of Eq.~(\ref{e:evolsystem}).  
Systems with this general form include most of the evolution
equations of interest in mathematical physics, including for example
the Einstein evolution equations, the Maxwell equations, the
incompressible fluid equations, etc.

In order to explore and test some of the ideas for controlling the
growth of constraints in these hyperbolic evolution systems, we adopt
a simple example system on which to perform our analysis and to carry
out numerical tests.  We have chosen to use a form of the vacuum
Maxwell evolution equations (introduced independently by
Kidder~\cite{Kidder2002} and Reula~\cite{Reula2002})
that fits nicely into this framework, and that admits constraint
violations if nothing is done to control them.  The dynamical fields
in this formulation are a co-vector that represents the electric field
$E_i$, and a second rank tensor $D_{ij}$ that represents the gradient
of the spatial parts of the vector potential
(i.e. $D_{ij}=\partial_iA_j$, although we impose the relationship
between $D_{ij}$ and the vector potential only implicitly as a
constraint on this system).  We refer to this as the fat Maxwell
system, since the usual representation of the Maxwell equations with
six dynamical field components is replaced with this larger
representation that has twelve.  The evolution equations for this
system are,
\begin{eqnarray} \partial_t E_i &=&
g^{ab}\partial_a(D_{ib}-D_{bi}),\label{e:eeq}\\ 
\partial_t D_{ij}&=&-\partial_iE_j-\partial_i\partial_j\phi,\label{e:deq} 
\end{eqnarray}
where $g_{ab}$ is the Euclidean metric with inverse $g^{ab}$, and
$\partial_a$ is the covariant derivative compatible with this metric,
(i.e.  just partial derivatives in Cartesian coordinates).  
The scalar potential $\phi$ is a gauge quantity assumed here
to be a given function of space and
time. This
system has the same general form as Eq.~\eqref{e:evolsystem} with
$u^\alpha=\{E_i,D_{ij}\}$.  In order to represent the vacuum
(i.e. charge and current free) Maxwell system these equations are also
subject to the constraints, ${\cal C}={\cal C}_{ijk}=0$, where
\begin{eqnarray}
{\cal C}&\equiv & g^{ab}\partial_a E_b,\label{e:diveconst}\\
{\cal C}_{ijk}&\equiv& \half (\partial_i D_{jk}-\partial_j D_{ik}).
\label{e:curlconst}
\end{eqnarray}
These constraints have the same general form as those described in
Eq.~\eqref{e:constraintdefs} with $c^A=\{{\cal C}, {\cal C}_{ijk}\}$.
The second of these constraints is the integrability condition that
guarantees that $D_{ij}$ is the gradient of a vector potential.  As
mentioned above we are free to add multiples of the constraints to
the evolution system:
\begin{eqnarray}
\partial_t E_i &=& g^{ab}\partial_a(D_{ib}-D_{bi})
+\gamma_{1} g^{ab}{\cal C}_{iab},\qquad\label{e:ewithceq}\\
\partial_t D_{ij} &=& -\partial_iE_j-\partial_i\partial_j\phi
+\gamma_{2} g_{ij}{\cal C},\label{e:dwithceq}
\end{eqnarray}
where $\gamma_{1}$ and $\gamma_{2}$ are constants.  The addition of
these constraint terms leaves the physical (constraint preserving)
solutions to the system unchanged, and also leaves the system with the
same basic structure as Eq.~\eqref{e:evolsystem}.

For hyperbolic evolution systems, such as those in
Eq.~\eqref{e:evolsystem}, it is often quite useful to decompose the
dynamical fields $u^\alpha$ into characteristic fields.  These
characteristic fields are defined with respect to a spatial direction
at each point, represented here by the unit normal co-vector field
$n_k$.  Given a direction field $n_k$ we define the eigenvectors
$e^{\hat \alpha}{}_\alpha$ of the characteristic matrix
$A^{k\,\alpha}{}_\beta$:
\begin{eqnarray}
e^{\hat\alpha}{}_\alpha n_k A^{k\,\alpha}{}_\beta=v_{(\hat\alpha)} 
e^{\hat\alpha}{}_\beta,\label{e:eigenvalueeq}
\end{eqnarray}
where $v_{(\hat \alpha)}$ denotes the eigenvalue (also called the
characteristic speed).  The index $\hat \alpha$ labels the various
eigenvectors and eigenvalues, and there is no summation over
$\hat\alpha$ in Eq.~\eqref{e:eigenvalueeq}.  Since we are interested
in hyperbolic evolution systems, the space of eigenvectors will have
the same dimension as the space of dynamical fields, and the matrix
$e^{\hat\alpha}{}_\beta$ will be invertible.  Given these
characteristic eigenvectors it is often useful to re-express the
dynamical fields in terms of this eigenvector basis.  Thus we define
the characteristic fields $u^{\hat\alpha}$ (or the characteristic
projection of the dynamical fields) as
\begin{eqnarray}
u^{\hat\alpha} = e^{\hat\alpha}{}_{\beta}u^\beta.
\label{e:characteristicparts}
\end{eqnarray}
It is straightforward to show that the evolution of the characteristic
fields is determined by
\begin{eqnarray}
\partial_tu^{\hat\alpha}+v_{(\hat\alpha)}n^k\partial_ku^{\hat\alpha}
&=&-e^{\hat \alpha}{}_\alpha P^{n}{}_kA^{k\,\alpha}{}_\beta\partial_n u^\beta
+e^{\hat\alpha}{}_\alpha F^\alpha\quad\nonumber\\
&&+\bigl(\partial_t e^{\hat\alpha}{}_{\alpha}+
v_{(\hat\alpha)}n^k\partial_k e^{\hat\alpha}{}_{\alpha}\bigr) u^\alpha,
\label{e:charevol}
\end{eqnarray}
where the projection operator orthogonal to $n_i$ is defined by
$P_{ij}=g_{ij}-n_in_j$, and spatial indices are raised and lowered
with $g^{ij}$ and $g_{ij}$ respectively.  

The characteristic fields for the fat Maxwell evolution system are
a collection of fields of the form $u^{\hat\alpha}=\{Z^1,Z^2_i,Z^3_{ij},
U^{1\pm}_i,U^{2\pm}\}$, where
\begin{eqnarray}
Z^1&=&2\gamma_{2}n^mn^n D_{mn}-(\gamma_{2}-1)P^{mn}D_{mn},
\label{e:Z1}\\
Z^2_i&=& P^m{}_in^n D_{mn},\\
Z^3_{ij}&=&\bigl(P^m{}_iP^n{}_j-\half P_{ij}P^{mn}\bigr)D_{mn},\\
U^{1\pm}_i&=&P^m{}_iE_m \pm n^mP^n{}_iD_{mn}
\pm \half(\gamma_{1}-2)P^m{}_in^nD_{mn},\label{e:U1}\nonumber\\
&&\\
U^{2\pm}&=&\pm n^mE_m-\half{\sqrt{\gamma_{1}\gamma_{2}}\over
\gamma_{2}}P^{mn}D_{mn}.\label{e:U2}
\end{eqnarray}
The characteristic fields $Z^1$, $Z^2_i$ and $Z^3_{ij}$ have
characteristic speed $0$; the fields $U^{1\pm}_i$ have speeds $\pm 1$,
and the fields $U^{2\pm}$ have speeds
$\pm\sqrt{\gamma_{1}\gamma_{2}}$.  All the characteristic speeds are
real, and the characteristic fields are linearly independent (and
depend continuously on the unit vector $n_k$) whenever
$\gamma_{1}\gamma_{2}> 0$.  Consequently the fat Maxwell
system is strongly hyperbolic when $\gamma_{1}\gamma_{2}> 0$.  We also
find that the fat Maxwell evolution system is symmetric hyperbolic when
the parameters $\gamma_{1}$ and $\gamma_{2}$ satisfy the more
restrictive conditions, $0<\gamma_{1}<4$, and $\third < \gamma_{2}$.

We note that the characteristic eigenvectors $e^{\hat\alpha}{}_\alpha$
for the fat Maxwell system depend only on the spatial metric $g_{ij}$ and
the normal vector $n_i$.  It follows that the last term on the right
side of Eq.~\eqref{e:charevol} does not depend on any derivatives of
$u^\alpha$ at all.  Thus the right side of Eq.~\eqref{e:charevol}
depends only on the transverse (to $n_i$) derivatives of $u^\alpha$:
\begin{eqnarray}
\partial_tu^{\hat\alpha}+v_{(\hat\alpha)}n^k\partial_ku^{\hat\alpha}
= G^{\hat\alpha}(u^\beta,P^k{}_n\partial_k u^\beta).
\end{eqnarray}
This important feature of the characteristic evolution equations
is satisfied by many systems of interest to us,
including the Einstein evolution system.

It is also useful to have the inverse transformation
$u^\alpha=e^\alpha{}_{\hat\alpha}u^{\hat\alpha}$, where
$e^\alpha{}_{\hat\alpha}$ is the inverse of $e^{\hat\alpha}{}_\alpha$.
For the fat Maxwell system this inverse transformation
has the form:
\begin{eqnarray}
E_i&=&\half(U^{1+}_i+U^{1-}_i)+\half n_i(U^{2+}-U^{2-}),\label{e:echar}\\
D_{ij}&=&\half n_in_j
\biggl[{Z^1\over \gamma_{2}}-{\gamma_{2}-1
\over\sqrt{\gamma_{1}\gamma_{2}}}
\bigl(U^{2+}+U^{2-}\bigr)\biggr]
\nonumber\\
&&-\half P_{ij} {\gamma_{2}
\over\sqrt{\gamma_{1}\gamma_{2}}}\bigl(U^{2+}+U^{2-}\bigr)
+Z^2_in_j\nonumber\\
&&+\half n_i\bigl[U^{1+}_j-U^{1-}_j-(\gamma_{1}-2)Z^2_j\bigr]
+Z^3_{ij}.\label{e:dchar}
\end{eqnarray}
The characteristic decomposition of the dynamical fields is essential
for fixing boundary conditions.  We will return to a more complete
discussion of boundary conditions in Sec.~\ref{s:boundaryconditions}.

\section{Active Constraint Control}
\label{s:constraintcontrol}

Unless the constraint evolution system Eq.~\eqref{e:constsystem} is
hyperbolic, it will not guarantee that the constraints remain
satisfied (within the domain of dependence of an initial surface) even
if they are satisfied initially.  Thus the constraint evolution system
must be hyperbolic in any self-consistent and physically reasonable
system of constrained evolution equations.  We assume that any system
considered here has a hyperbolic constraint evolution system.  We also
assume that the constraint evolution system satisfies the somewhat
stronger condition of symmetric hyperbolicity: In particular we assume
that there exists a symmetric, positive-definite tensor $S_{AB}$ on
the space of constraint fields which symmetrizes the characteristic
matrices of the constraint system,
\begin{eqnarray} S_{AC}A^{k\,C}{}_B \equiv A^k{}_{AB}=A^k{}_{BA}, 
\end{eqnarray} 
for all $k$.  When such a symmetrizer exists, we can define a natural
norm on the constraints: The constraint energy ${\cal E}$ and its
associated current ${\cal E}^k$ are defined by
\begin{eqnarray}
{\cal E}&=&S_{AB}\,c^Ac^B,\\
{\cal E}^k&=&A^k{}_{AB}\,c^Ac^B.
\end{eqnarray}
This constraint energy can be used to define a norm $\langle{\cal E}\hs\rangle$
on the constraints,
\begin{eqnarray}
\langle{\cal E}\hs\rangle=\int {\cal E} \,d^{\,3}x,\label{e:energynorm}
\end{eqnarray}
since $\langle{\cal E}\hs\rangle=0$ if and only if all the
constraints are satisfied at each point.  It is straightforward to
determine the time evolution of ${\cal E}$ using the constraint evolution
equations for any symmetric hyperbolic constraint evolution system:
\begin{eqnarray}
\partial_t{\cal E}+\partial_k{\cal E}^k = {\cal E}_{AB}c^Ac^B.
\label{e:constraintenergysystem}
\end{eqnarray}
The quantities ${\cal E}^k$ and ${\cal E}_{AB}$ may depend on the
dynamical fields $u^\alpha$ and their spatial derivatives $\partial_k
u^\alpha$ (but not on higher spatial derivatives of $u^\alpha$).

In an evolution system Eq.~\eqref{e:evolsystem} that
includes parameters $\gamma_a$ multiplying constraint terms, such as
the system defined by Eqs.~\eqref{e:ewithceq} and \eqref{e:dwithceq},
the associated constraint evolution system Eq.~\eqref{e:constsystem}
and the constraint energy system Eq.~\eqref{e:constraintenergysystem}
will also include terms that depend linearly on these parameters.
Integrating Eq.~\eqref{e:constraintenergysystem} over the spatial
slice for such a system, we get an expression for the time evolution
of the constraint norm which has the general form,
\begin{eqnarray}
\partial_t\langle{\cal E}\hs\rangle = Q + \gamma_a R^a,
\label{e:cnormeq}
\end{eqnarray}
where $Q$ and $R^a$ are integrals of quantities that depend on the
dynamical fields and their first spatial derivatives.  The basic idea
of active constraint control then is to adjust the parameters
$\gamma_a$ that appear in Eq.~\eqref{e:cnormeq} to control the
evolution of the constraint norm $\langle{\cal E}\hs\rangle$.  For
example the growth of $\langle{\cal E}\hs\rangle$ might be prevented
by making the right side of Eq.~\eqref{e:cnormeq} non-positive at the
beginning of each time step in the numerical evolution.  This control
mechanism is a special case of the constraint control method developed 
by Tiglio and
his collaborators~\cite{Tiglio2003a,Tiglio2003b}.  It differs from
Tiglio's particular implementation~\cite{Tiglio2003a,Tiglio2003b} in
that the quantities $Q$ and $R^a$ in our expression do not depend on
second derivatives of the dynamical fields.  Since these higher
derivatives are more difficult to evaluate accurately in a numerical
simulation, we expect that our constraint control mechanism will be
more stable and robust.

The constraints associated with the vacuum fat Maxwell system introduced
in Sec.~\ref{s:maxwell} satisfy the following evolution system as a
consequence of Eqs.~\eqref{e:ewithceq} and \eqref{e:dwithceq},
\begin{eqnarray}
\partial_t {\cal C} &=&
\gamma_{1} g^{ij}g^{ab} \partial_i {\cal C}_{jab},
\label{e:const1eq}\\
\partial_t {\cal C}_{ijk}&=& \half \gamma_{2} 
\bigl(g_{jk}\partial_i {\cal C} - g_{ik} \partial_j {\cal C}\bigr).
\label{e:const2eq}
\end{eqnarray}
This system has the same general form as Eq.~\eqref{e:constsystem}
with $c^A=\{{\cal C}, {\cal C}_{ijk}\}$.  In order to define a
constraint energy, we need this constraint evolution system to be
symmetric hyperbolic.  The most general symmetrizer for this system
(that can be constructed from the spatial metric $g_{ab}$) is given by
\begin{eqnarray}
dS^2&\equiv& S_{AB}dc^Adc^B\nonumber\\
&=&\chi_{1}g^{ij}g^{ab} d{\cal C}_{ij}d{\cal C}_{ab}
+\chi_{2}g^{ia}g^{jb}d\tilde {\cal C}_{ij}d\tilde {\cal C}_{ab}\nonumber\\
&&+\chi_{3}g^{ia}g^{jb}d{\cal C}_{[ij]}d{\cal C}_{[ab]}
+2\chi_3 {\gamma_2\over \gamma_1} d{\cal C}^2,
\label{eq:ConstraintSymmetrizer}
\end{eqnarray}
where 
\begin{eqnarray}
d{\cal C}_{ij}&=&g_{ic}\epsilon^{cab}d{\cal C}_{abj},\\
d\tilde {\cal C}_{ij}&=&
\half\bigl(\delta_{i}^{a}\delta_{j}^{b}
+\delta_{j}^{a}\delta_{i}^{b} -\twothirds
g_{ij}g^{ab}\bigr)d{\cal C}_{ab},
\end{eqnarray}
and $\epsilon^{ijk}$ is the totally antisymmetric tensor volume
element.  The parameters $\chi_{a}$ must be positive $\chi_{a}>0$, and
$\gamma_1\gamma_2$ must also be positive $\gamma_1\gamma_2>0$ to
ensure that $S_{AB}$ is positive definite.  We note that these
conditions put no additional limits on the allowed ranges of the
parameters: every strongly hyperbolic representation of the principal
evolution system has a symmetric hyperbolic constraint evolution
system.

We now evaluate the various quantities that determine the evolution of
the constraint energy, Eq.~\eqref{e:constraintenergysystem}, for the fat
Maxwell system.  We find:
\begin{eqnarray}
{\cal E}^k&=&-4\gamma_2\chi_3 {\cal C} g^{ij}g^{ka}{\cal C}_{aij},\\
{\cal E}_{AB}&=&0.
\end{eqnarray}
Thus the expression for the time derivative of the constraint energy
becomes,
\begin{equation}
\partial_t {\cal E} = 4\gamma_{2}\chi_{3} \partial_k
\bigl({\cal C} g^{ij} g^{ka}{\cal C}_{aij}\bigr).\label{e:maxeevol}
\end{equation}
The right side of Eq.~\eqref{e:maxeevol} is a divergence, so the
integral of this equation over a spatial surface results in an
expression that involves only boundary integrals:
\begin{equation}
\partial_t\langle {\cal E}\hs\rangle = 4\gamma_{2}\chi_{3}\oint  
{\cal C} g^{ij} n^{k}{\cal C}_{kij}\, d^{\,2}x,
\label{e:denergynorm}
\end{equation}
where $n^k$ is the outward directed unit normal to the boundary.
Active constraint control for this system consists then of adjusting
the sign of the parameter $\gamma_{2}$ to force the constraint norm
$\langle{\cal E}\hs\rangle$ to decrease with time whenever it gets
unacceptably large.

We note that the fat Maxwell system is rather degenerate, since the
right side of Eq.~\eqref{e:denergynorm} contains only a surface
integral.  Thus constraint violation in the fat Maxwell system arises
only from the influx of constraint violations through the timelike
boundaries of the computational domain.  This property makes this
system rather simpler than the Einstein evolution equations where
constraint violation can be generated from bulk terms in the equations
as well.  The simplicity of the fat Maxwell system allows us to study
how best to control the influx of constraint violations across
boundaries in some detail, but it does not allow us to evaluate how
effective these methods are for controlling violations that arise from
bulk terms in the equations.

\section{Constraint Preserving Boundary Conditions}
\label{s:boundaryconditions}

A standard boundary condition used for hyperbolic systems is the
maximally dissipative condition, which we define here to be the
condition that sets the incoming components of the dynamical fields to
zero.  (More generally the term maximally dissipative has been used to
describe a larger class of boundary conditions that guarantee that a
certain energy flux at the boundaries is strictly outgoing,
e.g. see~\cite{FriedrichNagy1999}.).  To impose such a condition, we
first decompose the dynamical fields into characteristic parts, as was
done in Eq.~\eqref{e:characteristicparts}, and then set to zero at the
boundary all those characteristic fields whose characteristic speeds
are negative. Let $\Pi^{\hat\alpha}{}_{\hat\beta}$ denote the
projection operator that annihilates all the non-incoming
characteristic fields: that is, let
\begin{eqnarray}
\Pi^{\hat\alpha}{}_{\hat\beta}u^{\hat\beta}=\left\{
\begin{array}{l}
u^{\hat\alpha} \quad \mbox{for}\quad v_{(\hat\alpha)}<0,\\
0 \quad\,\,\,\, \mbox{for}\quad v_{(\hat\alpha)}\geq 0.\\
\end{array}
\right.
\end{eqnarray}
For a maximally dissipative boundary condition, 
we set $\Pi^{\hat\alpha}{}_{\hat\beta}u^{\hat\beta}=0$ at the
boundaries. We often use a variation on this
boundary condition, in which we set to zero the time derivatives of
the incoming components of the characteristic fields:
\begin{eqnarray}
\Pi^{\hat\alpha}{}_{\hat\beta}\partial_t u^{\hat\beta}=0.
\end{eqnarray}
For the case of the fat Maxwell system discussed in Sec.~\ref{s:maxwell},
these ``freezing'' boundary conditions reduce to
\begin{eqnarray}
\partial_t U^{1-}_i=\partial_t U^{2-}=0,\label{e:maxwellfreezing}
\end{eqnarray}
where the incoming characteristic fields $U^{1-}_i$ and $U^{2-}$ are
defined in Eqs.~\eqref{e:U1} and \eqref{e:U2}.  As we shall see in
Sec.~\ref{s:numericalresults}, this ``freezing'' boundary condition
does a poor job of preventing the influx of constraint violations
through the boundaries.

Calabrese, {\it et al.}~\cite{Calabrese2002a} have proposed an
alternate method for constructing boundary conditions that prevent
the influx of constraint violations.  Their method involves
decomposing the constraint fields $c^A$ into characteristic parts:
\begin{eqnarray}
c^{\hat A}\equiv e^{\hat A}{}_B c^B,
\end{eqnarray}
where $e^{\hat A}{}_A$ represents the eigenvectors of the
characteristic matrix of the constraint evolution system,
\begin{eqnarray}
e^{\hat A}{}_B n_k A^{k\,B}{}_C=v_{(\hat A)} 
e^{\hat A}{}_C,\label{e:consteigenvalueeq}
\end{eqnarray}
and $v_{(\hat A)}$ represents the corresponding eigenvalue (or
characteristic speed).  The idea is to impose what amounts to
maximally dissipative boundary conditions on the constraint evolution
equations: that is, we set 
\begin{eqnarray}
\Pi^{\hat A}{}_{\hat B} c^{\hat B}=0,\label{e:constraintbc}
\end{eqnarray}
where $\Pi^{\hat A}{}_{\hat B}$ is the projection operator that
annihilates the non-incoming characteristic constraint fields.  This
condition must now be converted into a boundary condition on the
dynamical fields of the principal evolution system $u^\alpha$.  This
is done through the equation that defines the constraints in terms of
$u^\alpha$ and its derivatives, Eq.~\eqref{e:constraintdefs}. In
particular we solve Eq.~\eqref{e:constraintbc} for the normal
derivatives of the incoming characteristic fields, in terms of the
outgoing characteristic fields and tangential derivatives of the
incoming fields.  When this is possible, Eq.~\eqref{e:constraintbc}
becomes a Neumann-like boundary condition on (some of) the incoming
characteristic fields. This boundary condition has the following general form
\begin{eqnarray}
n^k\partial_k u^{\hat \beta}
=H^{\hat\alpha}[u^{\hat\beta},
(\delta^{\hat\beta}{}_{\hat\gamma}-\Pi^{\hat\beta}{}_{\hat\gamma})
\partial_ku^{\hat\gamma},
\Pi^{\hat\beta}{}_{\hat\gamma}P^k{}_n\partial_ku^{\hat\gamma}].\quad
\label{e:neumannbcgeneral}
\end{eqnarray}
We illustrate this procedure below more explicitly (and perhaps more
clearly) for the specific case of the fat Maxwell system.

The characteristic fields for the fat Maxwell constraint system are the
collection of fields of the form $c^{\hat A}=
\{Z^4_i,Z^5_{ij},U^{3\pm}\}$, where
\begin{eqnarray}
Z^4_i&=&{\cal C}_{[ik]}n^k,\label{e:Z4}\\
Z^5_{ij}&=&{\cal C}_{(ij)},\\
U^{3\pm}&=&-{\sqrt{\gamma_1\gamma_2}\over\gamma_1}{\cal C}
\pm n^kg^{ij}{\cal C}_{kij}.\label{e:U3}
\end{eqnarray}
The fields $Z^4_i$ and $Z^5_{ij}$ have characteristic speed 0, while
the fields $U^{3\pm}$ have speeds $\pm\sqrt{\gamma_1\gamma_2}$.  The
only incoming characteristic field is $U^{3-}$.  So the constraint
preserving boundary condition sets $U^{3-}=0$ on the boundaries of
the computational domain.  Using the definition of $U^{3-}$ above,
we see that this boundary condition is equivalent to the condition
\begin{eqnarray}
n^kg^{ij}{\cal C}_{kij}=
-{\sqrt{\gamma_1\gamma_2}\over\gamma_1}{\cal C}
\label{e:constraintbc0}
\end{eqnarray}
on the boundaries.  For a solution that satisfies the constraint
preserving boundary condition, Eq.~\eqref{e:constraintbc0},
the evolution of the constraint energy
norm Eq.~\eqref{e:denergynorm} becomes
\begin{equation}
\partial_t\langle {\cal E}\hs\rangle =-4\chi_{3}\sqrt{\gamma_1\gamma_2}
{\gamma_{2}\over\gamma_1}\oint  {\cal C}^2 \,d^{\,2}x\le0.
\label{e:denergynorm2}
\end{equation}
Thus the constraint preserving boundary condition ensures that
the constraint norm does not grow.  Quite generally the constraint
preserving boundary conditions of this type will ensure that
surface flux terms do not contribute to the growth of the
constraint energy. 

In order to convert the constraint preserving boundary condition into
an explicit condition on the dynamical fields, we must express the
incoming constraint field $U^{3-}$ in terms of the characteristic
fields $u^{\hat\alpha}$.  Using Eqs.~\eqref{e:Z1}--\eqref{e:U2} and
\eqref{e:U3} we obtain 
\begin{eqnarray}
U^{3-}&=&{\sqrt{\gamma_1\gamma_2}\over\gamma_1}\bigl[n^k\partial_k U^{2-}
-\half P^{ij}\partial_i(U^{1+}_j+U^{1-}_j)\bigr]\nonumber\\
&&+\fourth P^{ij}\partial_i\bigl[U^{1+}_j-U^{1-}_j-(\gamma_1-2)Z^2_j\bigr].
\end{eqnarray}
Setting $U^{3-}=0$ we obtain an expression for the normal derivative
of $U^{2-}$:
\begin{eqnarray}
&&n^k\partial_k U^{2-}
=\half P^{ij}\partial_i(U^{1+}_j+U^{1-}_j)\nonumber\\
&&\qquad-\fourth{\gamma_1\over\sqrt{\gamma_1\gamma_2}} 
P^{ij}\partial_i\bigl[U^{1+}_j-U^{1-}_j-(\gamma_1-2)Z^2_j\bigr].\qquad
\label{e:neumannbc}
\end{eqnarray}
This has the form of a Neumann-like boundary condition on $U^{2-}$, and has
the same form as the general expression
Eq.~\eqref{e:neumannbcgeneral}.  

The version of our code used to perform the numerical tests
described in Sec.~\ref{s:numericalresults} imposes boundary conditions
on the time derivatives of the incoming characteristic fields.  We
therefore convert the Neumann-like boundary condition on $U^{2-}$ in
Eq.~\eqref{e:neumannbc} into a condition on its time derivative using
the characteristic field evolution equation, Eq.~\eqref{e:charevol}.
We simply replace the normal derivative $n^k\partial_k U^{2-}$ that
appears in Eq.~\eqref{e:charevol} with the expression from
Eq.~\eqref{e:neumannbc}.  Simplifying the results gives the
following equation for the time derivative of $U^{2-}$ at the
boundary,
\begin{eqnarray}
\partial_t U^{2-} &=&\half{\sqrt{\gamma_1\gamma_2}\over\gamma_1}
P^{ij}(\partial_iE_j+\partial_i\partial_j\phi)
\nonumber \\ &&
+2P^{ij}n^k\partial_i D_{[jk]}.
\label{eq:ConstraintBc}
\end{eqnarray}
This condition together with the freezing boundary conditions
$\partial_t U^{1-}_i$ on the remaining incoming characteristic fields
constitute our version of constraint preserving boundary conditions
on the fat Maxwell system.

\section{Numerical Results}
\label{s:numericalresults}

In this section we present numerical experiments that illustrate the
effectiveness of the various constraint control strategies discussed
in this paper.  All of these tests use the fat Maxwell evolution
system, with a variety of topologies for the computational domain and
with a variety of initial data for the dynamical fields.  In
Sec.~\ref{s:nocontrol} we illustrate what happens when the equations
are solved without any constraint control. These tests show that
significant constraint violations (and in some cases constraint
violating instabilities) occur in dynamical solutions of the fat
Maxwell system on computational domains with open boundaries.  In
Sec.~\ref{s:activecontrol} we study the use of the active constraint
control mechanism described in Sec.~\ref{s:constraintcontrol}.  Our
tests show that this method is not numerically convergent, and is not
very effective in controlling the growth of constraints in this
system.  And finally in Sec.~\ref{s:bccontrol} we describe the results
of using the constraint preserving boundary conditions described in
Sec.~\ref{s:boundaryconditions}.  This method is shown to be
numerically convergent and quite effective in controlling the growth of
constraints in the symmetric hyperbolic subset of the fat Maxwell system.

All numerical computations presented here are performed using a
pseudospectral collocation method. Our numerical methods are
essentially the same as those we have applied to the evolution problem
in full general
relativity~\cite{Kidder2000a,Kidder2001,Lindblom2002,Scheel2002} and
in studies of scalar fields in Kerr spacetime~\cite{Scheel2004}.
Given a system of partial differential equations
\begin{equation} 
\partial_t u^\alpha(\mathbf{x},t) = 
{\cal F}^\alpha[u(\mathbf{x},t),\partial_i u(\mathbf{x},t) ],
\label{diffeq}
\end{equation}
where $u^\alpha$ is a vector of dynamical fields, the solution
$u^\alpha(\mathbf{x},t)$ is expressed as a time-dependent linear
combination of $N$ spatial basis functions $\phi_k(\mathbf{x})$:
\begin{equation}
u^\alpha_N(\mathbf{x},t) = 
        \sum_{k=0}^{N-1}\tilde{u}^\alpha_k(t) \phi_k(\mathbf{x}).
\label{decom}
\end{equation}
Spatial derivatives are evaluated analytically using the known
derivatives of the basis functions:
\begin{equation}
\partial_i u^\alpha_N(\mathbf{x},t) 
= \sum_{k=0}^{N-1}\tilde{u}^\alpha_k(t)
  \partial_i\phi_k(\mathbf{x}).
\label{decomderiv}
\end{equation}
The coefficients $\tilde{u}^\alpha_k(t)$ are chosen so that
Eq.~(\ref{diffeq}) is satisfied exactly at $N_c$ collocation points
$\mathbf{x}_i$ selected from the spatial domain.  The values of the
coefficients are obtained by the inverse transform
\begin{equation}
\tilde{u}^\alpha_k(t) = \sum_{i=0}^{N_c-1}u^\alpha_N(\mathbf{x}_i,t)
                       \phi_k(\mathbf{x}_i) w_i, 
\label{invdecom}
\end{equation}
where $w_i$ are weights specific to the choice of basis functions and
collocation points. It is straightforward to transform between the
spectral coefficients $\tilde{u}^\alpha_k(t)$ and the function values
at the collocation points $u^\alpha_N(\mathbf{x}_i,t)$ using
Eqs.~(\ref{decom}) and (\ref{invdecom}).  The partial differential
equations, Eq.~(\ref{diffeq}), are now rewritten using
Eqs.~(\ref{decom})--(\ref{invdecom}) as a set of {\it ordinary\/}
differential equations for the function values at the collocation
points,
\begin{equation} 
\partial_t u^\alpha_N(\mathbf{x}_i,t) 
                      = {\cal G}^\alpha_i [u_N(\mathbf{x}_j,t)],
\label{odiffeq}
\end{equation}
where ${\cal G}^\alpha_i$ depends on $u^\alpha_N(\mathbf{x}_j,t)$ for
all $j$.  This system of ordinary differential equations,
Eq.~(\ref{odiffeq}), is integrated in time using a fourth-order
Runge-Kutta method.  Boundary conditions are incorporated into the
right-hand side of Eq.~(\ref{odiffeq}) using the technique of
Bj{\o}rhus~\cite{Bjorhus1995}. The time
step is typically chosen to be half the distance
between the closest collocation points, which ensures that the
Courant condition is satisfied.

In order to provide a quantitative measure of convergence and the
amount of constraint violation of our numerical solutions, we have
defined several norms on the constraints $c^A$ and the dynamical
fields $u^\alpha$.  We have already defined the constraint energy
$\langle{\cal E}\hs\rangle$ in Eq.~(\ref{e:energynorm}), which provides a
norm on the constraint fields.  In computing $\langle {\cal E} \hs\rangle$
for these numerical studies we fix $\chi_1=\chi_2=\chi_3=1$.  We also
define norms on the dynamical fields themselves:
\begin{eqnarray}
||u||^2_{L^2} &\equiv&        \int (E_i E^i + D_{ij} D^{ij})\,d^{\,3}x , 
\label{eq:L2NormDef} \\
||u||_{L{}^\infty}^2 &\equiv& \max(E_i E^i + D_{ij} D^{ij}).
\label{eq:SupNormDef} 
\end{eqnarray}
We compute the volume integrals in these norms, {\it e.g.} in
Eq.~(\ref{e:energynorm}) or (\ref{eq:L2NormDef}), exactly using the
appropriate form of Gaussian quadrature, and the maximum in
Eq.~(\ref{eq:SupNormDef}) is taken over the appropriate set of
collocation points at a particular instant of time.  These norms are
most useful for comparing solutions evaluated with different numerical
resolutions.  Thus we define
\begin{eqnarray}
||\delta u||^2_{L^2} &\equiv&  \int (\delta E_i \delta E^i 
                                  + \delta D_{ij} \delta D^{ij})\,d^{\,3}x , 
\label{eq:L2DiffNormDef} \\
||\delta u||_{L^\infty}^2 &\equiv& \max
                     (\delta E_i \delta E^i + \delta D_{ij} \delta D^{ij}),
\label{eq:SupDiffNormDef}
\end{eqnarray}
where $\delta u^\alpha=\{\delta E_i,\delta D_{ij}\}$ is the difference
between $u^\alpha$ at a given resolution and $u^\alpha$ at the best
(highest) resolution we computed.  Differences between quantities at
different resolutions are computed by evaluating and then subtracting
the spectral series for each resolution at the points on the finer
grid.  In order to provide meaningful scales for these normed
quantities we typically plot dimensionless ratios of expressions such
as $||\delta u||^2_{L^2}/||u_{\rm best}||^2_{L^2}$ and $||\delta
u||_{L^\infty}^2/||u_{\rm best}||_{L^\infty}^2$.  In the case of the
constraint energy we typically plot $\langle{\cal E}\hs\rangle/||\partial u
||^2$, where
\begin{equation}
||\partial u||^2\equiv
\int (\partial_kE_i \partial^kE^i + \partial_kD_{ij} \partial^kD^{ij})
\,d^{\,3}x,
\end{equation}
is a norm on the gradients of the fields.  We are interested in seeing
how these ratios behave as the resolution of the numerical solution is
increased: Order unity values of these ratios, $||\delta
u||^2/||u_{\rm best}||^2$ or $\langle{\cal E}\hs\rangle/||\partial u
||^2$, indicate a complete lack of numerical convergence or solutions
that are dominated by constraint violations, respectively.  Values of
these ratios of order $10^{-34}$ correspond to double precision
roundoff error.

\subsection{No Constraint Control}
\label{s:nocontrol}

In this section we illustrate the results of finding numerical
solutions to the fat Maxwell evolution system
Eqs.~(\ref{e:ewithceq})--(\ref{e:dwithceq}) using no constraint
control at all.  We examine three separate cases: First we look at
evolutions on a computational domain with topology $T^3$, a 3-torus.
The differential equations governing the fat Maxwell system allow no
constraint growth on domains without boundaries.  So this first test
is to verify that our code accurately reproduces ``exact'' constraint
conservation in this case.  Next we examine the evolution of a
representation of the static Coulomb solution on a computational
domain with topology $S^2\times R$, a spherical shell.  Finally we
study a highly dynamical solution on a computational domain with
topology $S^2\times R$ using freezing boundary conditions and no
constraint control.

%
\begin{figure}
\begin{center}
\includegraphics[width=3.0in]{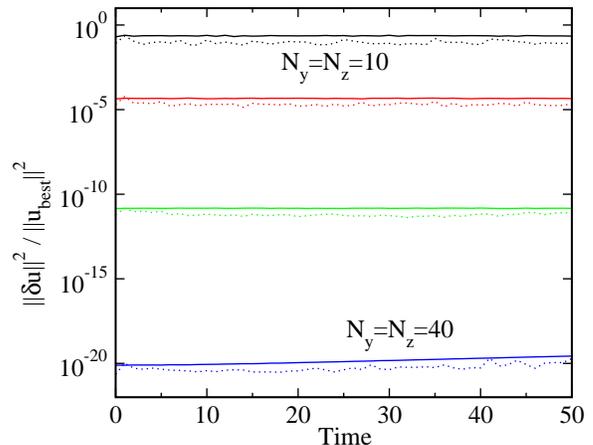}
\end{center}
\caption{Convergence test for fat Maxwell on $T^3$. Shown are norms of
differences between solutions at different resolutions: solid curves
use the $L^2$ norms while the dotted curves use the $L^\infty$ norms.}
\label{f:TorusConvergence}
\end{figure}
%
%
\begin{figure}
\begin{center}
\includegraphics[width=3.0in]{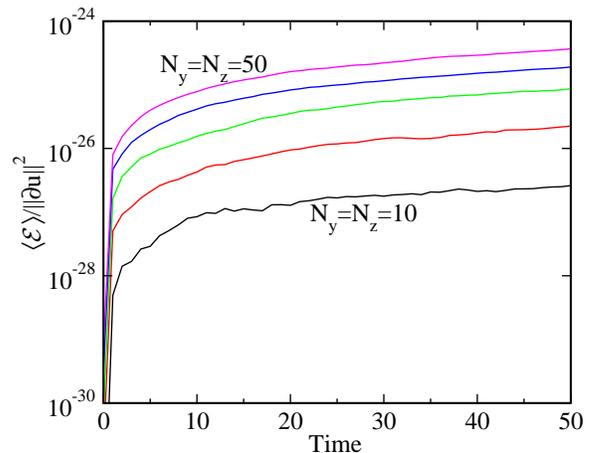}
\end{center}
\caption{Constraint violation for fat Maxwell on $T^3$. Shown is the
constraint energy $\langle {\cal E} \hs\rangle$ divided by the norm of the
derivatives of the fundamental variables.}
\label{f:TorusConstraint}
\end{figure}
%

The evolution of the constraint energy norm $\langle {\cal E}\hs\rangle$ for
the fat Maxwell system is driven entirely by a boundary term,
Eq.~(\ref{e:denergynorm}).  Thus we expect the constraints to be
satisfied exactly for evolutions on a computational domain without
boundary.  To confirm that our numerical code correctly models this,
we solve Eqs.~(\ref{e:ewithceq})--(\ref{e:dwithceq}) on a
computational domain with topology $T^3$, {\it i.e.} within a 3-torus.
In particular we choose coordinates $x$, $y$, and $z$ in the interval
$[0,2\pi]$, and impose periodic boundary conditions.  The basis
functions used in our pseudospectral method are sines and cosines. We
use initial data for this test in which the electric field is set to
zero, $E_i=0$, and each component of the vector potential is set to be
a cylindrical Gaussian pulse:
\begin{equation}
A_x=A_y=A_z = e^{-[(y-c_y)^2+(z-c_z)^2]/w^2},
\end{equation}
where the width of the pulse is set to $w=0.5$, and the center is
placed in the middle of the computational domain, $c_y=c_z=3.14$.  The
shape of this pulse is selected so that the value of the Gaussian
falls below double precision roundoff, $10^{-17}$, at the periodicity
``boundaries'' of the domain, $y=0$ and $y=2\pi$ etc.  This ensures
that these data are smooth on $T^3$ to roundoff accuracy.  The initial
data for $D_{ij}$ are set to the numerically determined values of
$\partial_iA_j$.  We use the gauge choice $\phi=0$ throughout this
evolution.  Because these initial data are effectively
two-dimensional, we can place as few as two collocation points in the
$x$ direction for computational efficiency.

Figure~\ref{f:TorusConvergence} shows a convergence plot for this case
that was run with evolution parameter values
$\gamma_1=1/\gamma_2=-0.1$, and resolutions $N_y=N_z=10, 20, 30, 40$, and $50$
collocation points.  We see from Fig.~\ref{f:TorusConvergence}
that the differences converge to zero as the resolution is increased.
Figure~\ref{f:TorusConstraint} illustrates the amount of constraint
violation in these runs.  These curves, which increase approximately
linearly with time, have magnitudes that are roughly proportional to
the number of numerical operations performed multiplied by double
precision roundoff error.  Thus, the finer resolutions have larger
errors than the coarser ones, since the finer resolutions require a
larger number of timesteps and a larger number of numerical operations
at each step.  As expected from Eq.~(\ref{e:denergynorm}), we see that
the constraints are satisfied essentially exactly when the domain has
no boundaries.  We have also computed evolutions for these initial
data on $T^3$ using other values of the evolution parameters.  In
particular we have computed evolutions with $\gamma_1=1/\gamma_2=0.1$,
and also evolutions that switch back and forth between these positive
and negative values at each time step.  In all of these cases, we find
the evolutions to be convergent with roundoff level constraint
violation.

Next we turn our attention to solving the evolution equations on a
computational domain with topology $S^2\times R$, {\it i.e.} within a
spherical shell.  For our basis functions we choose Chebyshev
polynomials for the radial coordinate and spherical harmonics for the
angular coordinates. Although our basis functions are written in
$(r,\theta,\varphi)$ coordinates, our fundamental variables are the
Cartesian components of $E_i$ and $D_{ij}$. To eliminate
high-frequency numerical instabilities that sometimes develop during
our simulations in $S^2\times R$, we apply a filter to the right-hand
sides of Eqs.~(\ref{odiffeq}) before and after incorporating boundary
conditions via the Bj{\o}rhus algorithm. The filter consists of simply
setting high-frequency spherical harmonic coefficients to zero: If
$\ell_{\rm max}$ is the largest index used in the basis functions
$Y_{\ell m}$ at a particular resolution, then the largest $\ell$
retained in the right-hand side of the $E_i$ equations after filtering
is $2\ell_{\rm max}/3-1$, and the largest $\ell$ retained in the
right-hand sides of the $D_{ij}$ equations is $2\ell_{\rm max}/3$.
This filter is a variation on the standard 2/3 rule used to remove the
inevitable effects of aliasing whenever functions are multiplied using
spectral methods~\cite{Boyd1989}.

%
\begin{figure}
\begin{center}
\includegraphics[width=3.0in]{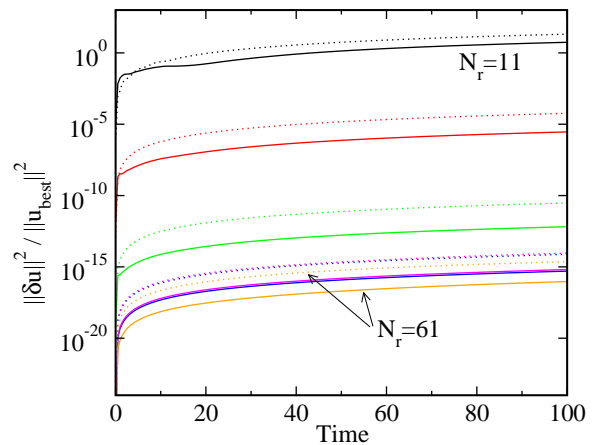}
\end{center}
\caption{Convergence test for fat Maxwell on $S^2\times R$ with static
point charge initial data.  Shown are norms of differences between
solutions at different resolutions: solid curves use $L^2$ norms
and dotted curves use $L^\infty$ norms.}
\label{f:SphereFreezingBcPtConvergence}
\end{figure}
%
\begin{figure}
\begin{center}
\includegraphics[width=3.0in]{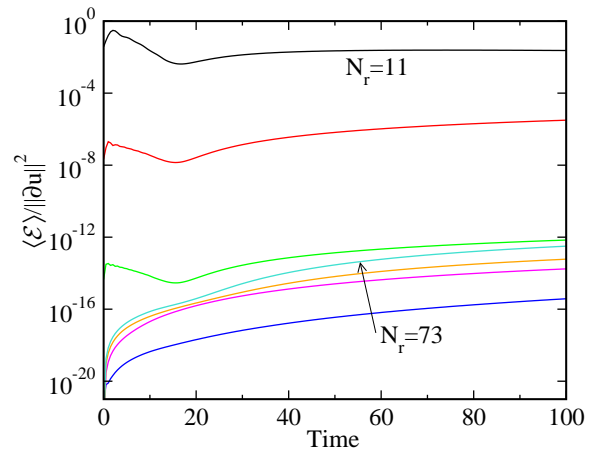}
\end{center}
\caption{Constraint violation for fat Maxwell on $S^2\times R$ with
static point charge initial data. Shown is the constraint energy
$\langle {\cal E} \hs\rangle$ divided by the norm of the derivatives of
the fundamental variables.}
\label{f:SphereFreezingBcPtConstraint}
\end{figure}

For the first of these tests we choose initial data
that corresponds to a static point charge that is located in the hole
in the center of the computational domain.  Thus we choose initial
data with $E_i=-\partial_i\phi =r^{-2}\partial_i r$ and $D_{ij}=0$,
appropriate for a unit point charge at rest at the origin.  We then
solve Eqs.~(\ref{e:ewithceq})--(\ref{e:dwithceq}) with
$\gamma_1=1/\gamma_2=0.1$ on a computational
domain with topology $S^2\times R$, defined by $1.9\leq r \leq 11.9$.
(This is the same computational domain that we typically use to evolve
single black hole spacetimes.)  At both the inner and outer spherical
boundaries we set the time derivatives of the incoming characteristic
fields to zero, i.e. we impose freezing boundary conditions,
Eq.~(\ref{e:maxwellfreezing}).  The scalar potential $\phi$ is held
constant in time. 
We find that these numerical
evolutions are stable and convergent and the constraints are preserved,
as shown in Figs.~\ref{f:SphereFreezingBcPtConvergence}
and~\ref{f:SphereFreezingBcPtConstraint}. 
These computations were performed with radial resolutions
$N_r=11,21,31,41,51,61,$ and $73$ collocation points, and a fixed
angular resolution with spherical harmonic index $\ell_{\rm max}=5$ (or
equivalently, $N_\theta=6$ and $N_\varphi=12$ angular collocation points). 
For $\ell_{\rm max}=9$ the results are indistinguishable on the
scale of Figs.~\ref{f:SphereFreezingBcPtConvergence}
and~\ref{f:SphereFreezingBcPtConstraint} except at the highest
radial resolutions, indicating
that the radial and temporal truncation errors dominate, as expected
for a solution with little angular structure.
This is a case (as we shall
see) in which a time-independent solution is not always the best test
problem to investigate the constraint-preserving properties of a
system of evolution equations.

Finally, we examine a highly dynamical solution of the fat Maxwell
system on the computational domain $S^2\times R$, defined by $1.9\leq
r \leq 11.9$.  For this solution we choose initial data with $E_i=0$
and
\begin{equation}
A_x=A_y=A_z = e^{-(r-r_0)^2/w^2},
\label{eq:SphericalGaussian}
\end{equation}
where $r_0=6.5$ and $w=1.0$.  The initial values of $D_{ij}$ are set
to the numerically determined values of $\partial_i A_j$.  These
initial data correspond to a pulse of radiation centered at $r=r_0$.
This pulse is neither spherically symmetric nor even axially
symmetric, because the Cartesian components of the vector potential
are set to spherically symmetric functions in
Eq.~(\ref{eq:SphericalGaussian}); however, only a small number
of spherical harmonics are sufficient to represent the full solution.
The scalar potential is set to
$\phi=0$ for these solutions, and we impose freezing boundary
conditions, Eq.~(\ref{e:maxwellfreezing}), on the incoming
characteristic fields.

%
\begin{figure}
\begin{center}
\includegraphics[width=3.0in]{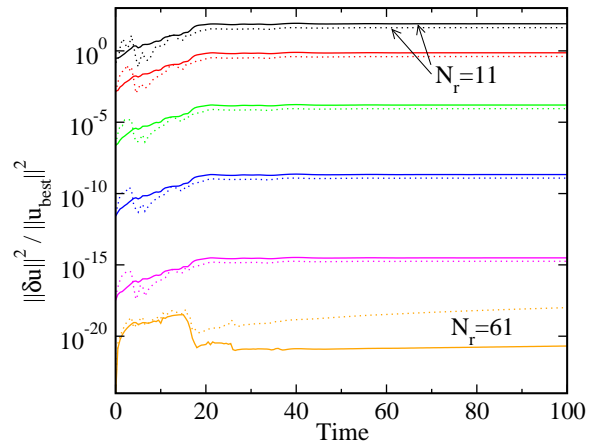}
\end{center}
\caption{Convergence test for a dynamical solution of fat Maxwell on
$S^2\times R$ using freezing boundary conditions and positive values
of $\gamma_a$.  Shown are norms of differences between solutions at
different resolutions: solid curves use $L^2$ norms and dotted curves
use $L^\infty$ norms.}
\label{f:SphereFreezingBcConvergence}
\end{figure}
%
\begin{figure}
\begin{center}
\includegraphics[width=3.0in]{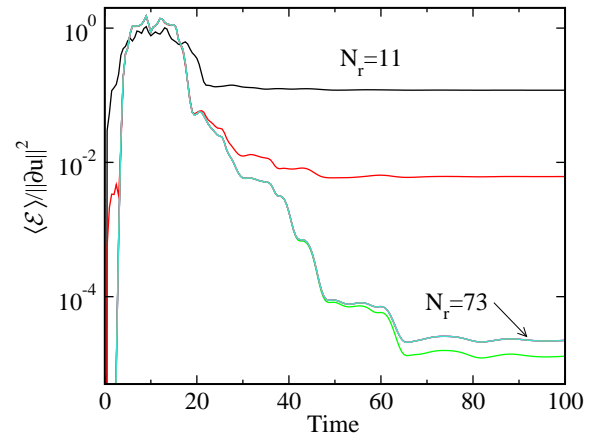}
\end{center}
\caption{Constraint violation for a dynamical solution of fat Maxwell
on $S^2\times R$ using freezing boundary conditions and positive
values of $\gamma_a$.  Shown is the constraint energy $\langle {\cal E}
\hs\rangle$ divided by the norm of the derivatives of the fundamental
variables.}
\label{f:SphereFreezingBcConstraint}
\end{figure}
%

Figure~\ref{f:SphereFreezingBcConvergence} shows a convergence plot
for this case, using evolution parameters $\gamma_1=1/\gamma_2=0.1$
and $\ell_{\rm max}=5$,
which confirms that the numerical solution is convergent.  
For $\ell_{\rm max}=9$ the results are indistinguishable on the
scale of the figure.
Figure~\ref{f:SphereFreezingBcConstraint}
shows that significant constraint violations do exist in these
solutions with seven different radial resolutions: the data from the
three highest resolutions coincide on the scale of this diagram.  Thus
the constraints are violated, but the constraint energy is still
convergent in these solutions.  This indicates that the constraint
violation is a property of the true solution of the continuum
equations with freezing boundary conditions, rather than an effect
caused by a defective numerical method.  The constraint violation
appears to be generated as the initially constraint-satisfying waves
interact with the boundaries, starting at about $t\approx 4$.

%
\begin{figure}
\begin{center}
\includegraphics[width=3.0in]{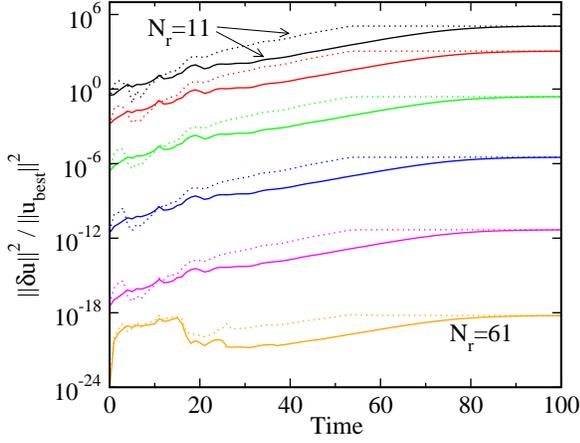}
\end{center}
\caption{Convergence test for fat Maxwell on $S^2\times R$ using
freezing boundary conditions and negative values of $\gamma_a$.  Shown
are norms of differences between solutions at different resolutions:
solid curves use $L^2$ norms and dotted curves use $L^\infty$ norms.}
\label{f:SphereFreezingBcMinusConvergence}
\end{figure}
%
\begin{figure}
\begin{center}
\includegraphics[width=3.0in]{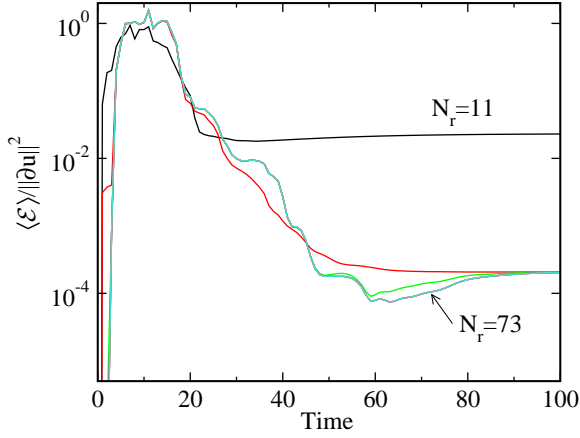}
\end{center}
\caption{Constraint violation for fat Maxwell on $S^2\times R$ using
freezing boundary conditions and negative values of $\gamma_a$.  Shown
is the ratio of the constraint energy to the norm of the derivatives
of the dynamical fields.}
\label{f:SphereFreezingBcMinusConstraint}
\end{figure}
%

\begin{figure}
\begin{center}
\includegraphics[width=3.0in]{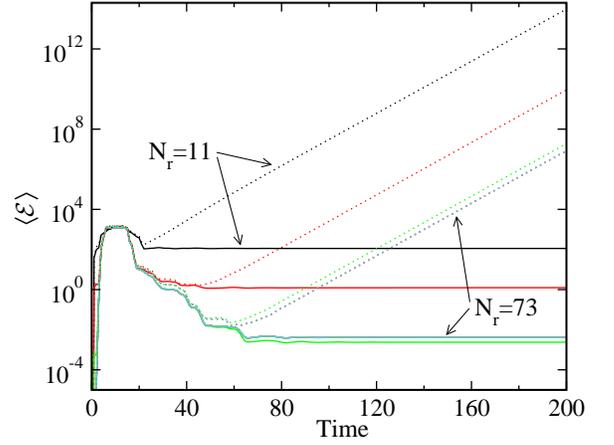}
\end{center}
\caption{Constraint violation for fat Maxwell on $S^2\times R$.  Solid
curves are for $\gamma_a>0$ while dotted curves are for $\gamma_a<0$
evolution parameters.  Seven different resolutions are depicted for
each sign of $\gamma_a$, but only the lowest resolution curves are
distinct at the scales shown.}
\label{f:SphereFreezingBcConstraints}
\end{figure}

We have also computed evolutions for these dynamical initial data
using negative values of the evolution parameters:
$\gamma_1=1/\gamma_2=-0.1$; the results are depicted in
Figs.~\ref{f:SphereFreezingBcMinusConvergence}
and~\ref{f:SphereFreezingBcMinusConstraint}. Since the product
$\gamma_1 \gamma_2$ is unchanged from the previous runs, the
characteristic speeds of the system remain the same. And the
definition of the constraint energy ${\cal E}$ (which depends on the
ratio $\gamma_1/\gamma_2$) also remains unchanged; so this allows us
to meaningfully compare ${\cal E}$ for the two cases.  For
$\gamma_a<0$ the fundamental evolution system,
Eqs.~(\ref{e:ewithceq})--(\ref{e:dwithceq}), is strongly but no longer
symmetric hyperbolic as in the $\gamma_a>0$ case.
Figures~\ref{f:SphereFreezingBcMinusConvergence}
and~\ref{f:SphereFreezingBcMinusConstraint} show that these evolutions
with negative values of $\gamma_a$ appear to be convergent, and have
fractional constraint violations that are comparable with those in the
positive $\gamma_a$ case.  However as illustrated in
Fig.~\ref{f:SphereFreezingBcConstraints}, the evolutions in the
negative $\gamma_a$ case have constraint violating instabilities.

%

\begin{figure}
\begin{center}
\includegraphics[width=3.0in]{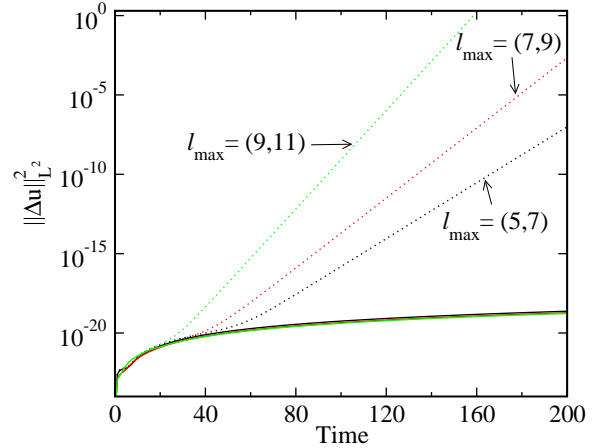}
\end{center}
\caption{Angular convergence test for a dynamical solution of fat
Maxwell, showing numerical instability for $\gamma_a<0$. Shown is a
norm similar to $||\delta u||_{L^2}$ defined in
Eq.~(\ref{eq:L2DiffNormDef}) except that the differences are taken
between quantities at two different angular
resolutions and fixed $N_r=73$.  Dotted curves show results for
$\gamma_1=1/\gamma_2=-0.1$, and are labeled by the two angular
resolutions that are subtracted.  Solid curves show that the same
quantities for the case of $\gamma_1=1/\gamma_2=+0.1$ are
convergent.}
\label{f:AngularConvergence}
\end{figure}

We note that we also find a {\it numerical instability\/} for the
$\gamma_a<0$ case, and apparently for all cases in which the
fundamental evolution system is strongly but not symmetric hyperbolic.
This numerical instability appears to be associated with the angular
discretization. It grows exponentially in time, and becomes worse at
higher angular resolutions.  However, for the angular resolutions we
use, and for choices of $\gamma_a$ near the symmetric hyperbolic
range, such as the case shown here, $\gamma_1=1/\gamma_2=-0.1$, the
numerical instability is negligible compared to the
constraint-violating instability shown in
Fig.~\ref{f:SphereFreezingBcConstraints}. Only by going to higher
angular resolution can one see any nonconvergent growth at all on the
time scales we consider here. For $\ell_{\rm max}=9$ the instability
is visible only at late times ($t\approx 200$) for the highest radial
resolutions in the quantities $||\delta u||^2_{L^2}$ and $||\delta
u||_{L^\infty}^2$, and is not visible in plots of $\langle {\cal E}
\hs\rangle$.  To construct a quantity that is sensitive to this
instability, we repeated the runs shown in
Figs.~\ref{f:SphereFreezingBcConvergence}--\ref{f:SphereFreezingBcConstraints}
at angular resolutions $\ell_{\rm max}=5$, $7$, $9$, and $11$ and
computed the norms of differences of the fundamental fields at
adjacent angular resolutions.  These norms are plotted in
Fig.~\ref{f:AngularConvergence}.

%

We see no indication of this numerical instability for values of
$\gamma_a$ in which the fundamental evolution system is symmetric
hyperbolic (for example, the solid curves in
Fig.~\ref{f:AngularConvergence} are convergent). For choices of
$\gamma_a$ very far from the symmetric hyperbolic range (such as
$\gamma_1=1/\gamma_2=10$), the instability grows much more rapidly and
dominates the results.  Although it is possible that the numerical
instability can be cured by modifying our angular filtering algorithm,
for the purpose of this paper we simply consider only values of
$\gamma_a$ and angular resolutions for which the timescale of this
instability is longer than that of other effects we wish to study.

\subsection{Active Constraint Control}
\label{s:activecontrol}

\begin{figure}
\begin{center}
\includegraphics[width=3.0in]{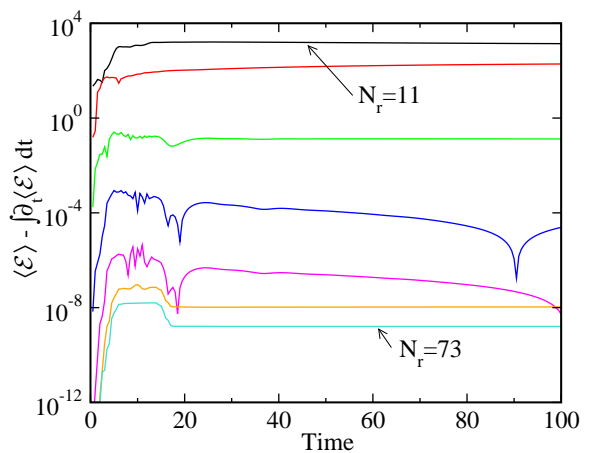}
\end{center}
\caption{Violation of the constraint energy evolution equation,
Eq.~(\ref{e:denergynorm}), for fat Maxwell on $S^2\times R$ with
freezing boundary conditions and $\gamma_a>0$. Plotted is the
difference between $\langle {\cal E} \hs\rangle$ and the time integral of
the right side of Eq.~(\ref{e:denergynorm}) for each resolution.}
\label{f:SphereFreezingBcBdryRelation}
\end{figure}

In this section we investigate the use of the active constraint
control methods described in Sec.~\ref{s:constraintcontrol}.  For the
case of the fat Maxwell system this active control consists of
switching the sign of the evolution parameters $\gamma_1$ and
$\gamma_2$ to ensure that the constraint energy $\langle {\cal
E}\hs\rangle$ does not increase.  In the previous section we presented
two sets of numerical evolutions without constraint control differing
only by the signs of $\gamma_1$ and $\gamma_2$.  The characteristic
speeds and the definition of the energy $\langle{\cal E}\hs\rangle$
were the same for these evolutions.  Both evolutions were convergent
on the time scale considered here ($t\leq 100$),
and the fractional constraint violation was convergent and similar in
these cases on the same time scale.  We now
investigate the possibility of switching between these two cases
during a single evolution as a means of reducing the constraint
violations. The strategy is to monitor the quantity on the right side
of Eq.~(\ref{e:denergynorm}), and to change the signs of $\gamma_1$
and $\gamma_2$ whenever necessary to keep the right side negative.
This should ensure that the constraint energy norm $\langle {\cal
E}\hs\rangle$ is always decreasing, so the constraints should remain
satisfied.  Note that this method should work only as long as our
numerical solution satisfies the equation governing the evolution of
the constraint energy, Eq.~(\ref{e:denergynorm}).
Figure~\ref{f:SphereFreezingBcBdryRelation} illustrates for the case
$\gamma_a>0$ that this equation does remain satisfied to truncation
error level for the runs discussed in Sec.~\ref{s:nocontrol};
the plot for $\gamma_a<0$ is similar.
Consequently we expected good results from this active control method.

\begin{figure}
\begin{center}
\includegraphics[width=3.0in]{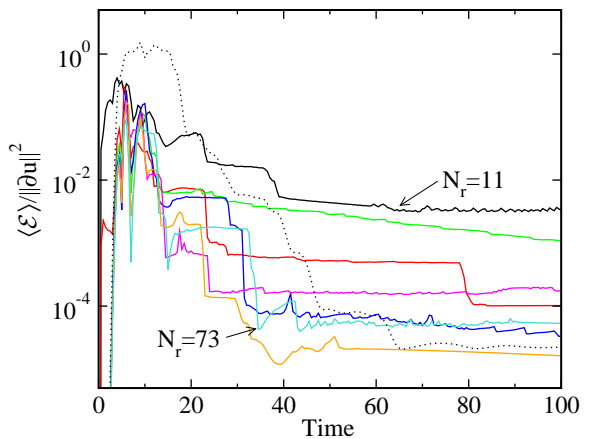}
\end{center}
\caption{Constraint violation (solid curves) for fat Maxwell on
$S^2\times R$ with active constraint control at every time step.
Dotted curve is the comparable uncontrolled case.}
\label{f:SphereControlConstraint}
\end{figure}

\begin{figure}
\begin{center}
\includegraphics[width=3.0in]{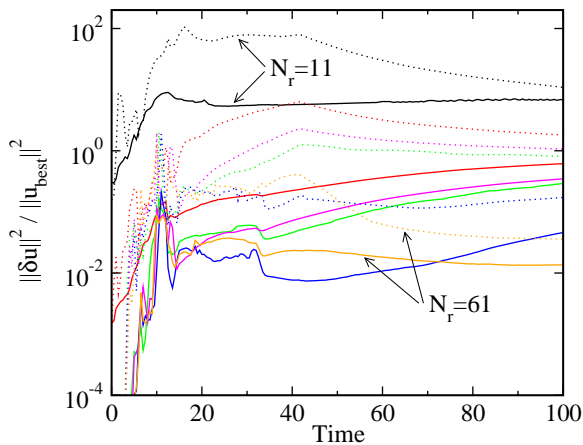}
\end{center}
\caption{Convergence test for fat Maxwell on $S^2\times R$,
for active constraint control at every time step.
Shown are norms of differences between solutions at different resolutions:
solid curves use $L^2$ norms and dotted curves use $L^\infty$ norms.}
\label{f:SphereControlConvergence}
\end{figure}

\begin{figure}
\begin{center}
\includegraphics[width=3.0in]{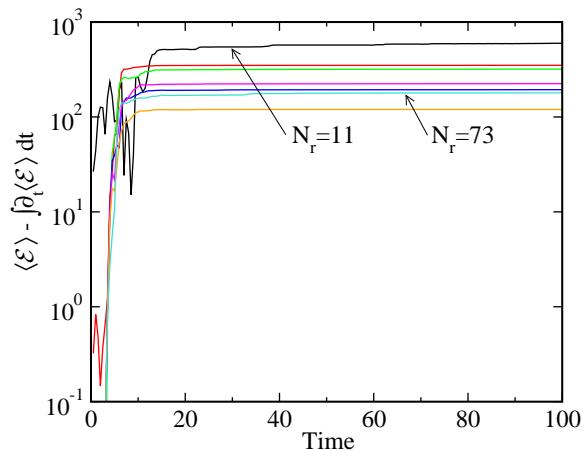}
\end{center}
\caption{Violation of the constraint energy evolution equation,
Eq.~(\ref{e:denergynorm}), for fat Maxwell on $S^2\times R$ with
with active constraint control at each time step. Plotted is the
difference between $\langle {\cal E} \hs\rangle$ and the time integral of
the right side of Eq.~(\ref{e:denergynorm}) for each resolution.}
\label{f:SphereControlBdryRelation}
\end{figure}

Figure~\ref{f:SphereControlConstraint} shows the constraint violation
for a case in which the $\gamma_a$ are allowed to change sign at every
time step in order to control the constraints. The signs are changed
only if the right side of Eq.~(\ref{e:denergynorm}) becomes positive,
and the current value of $\langle {\cal E}\hs\rangle$ exceeds the
value it had after the first timestep.  The latter condition is
intended to prevent sign changes that attempt to reduce the constraint
violation to less than truncation error.  Since this constraint
control method depends on Eq.~(\ref{e:denergynorm}) being satisfied,
control should only be possible to truncation error level at best.
Figure~\ref{f:SphereControlConstraint} shows that the maximum value of
the constraint is smaller than for the uncontrolled case (plotted as a
dotted line in Fig.~\ref{f:SphereControlConstraint}).  However, the
improvement is only an order of magnitude at best, even for the
highest resolution run.  Even more disturbing is the lack of
convergence of the constraint norm.  The fundamental fields do not
converge very rapidly (if at all) either, as can be seen from
Fig.~\ref{f:SphereControlConvergence}.  This lack of effective
constraint control is confirmed in
Fig.~\ref{f:SphereControlBdryRelation}, which shows that
Eq.~(\ref{e:denergynorm}) is not satisfied very well for this case.
It appears that the active constraint control mechanism severely
degrades the convergence of our numerical simulations in such a way
that Eq.~(\ref{e:denergynorm}) no longer holds to the needed or
expected accuracy.  Consequently the active constraint control method
is able to reduce the constraint violations only by a small amount
over the uncontrolled case.

%
%
%
%

\begin{figure}
\begin{center}
\includegraphics[width=3.0in]{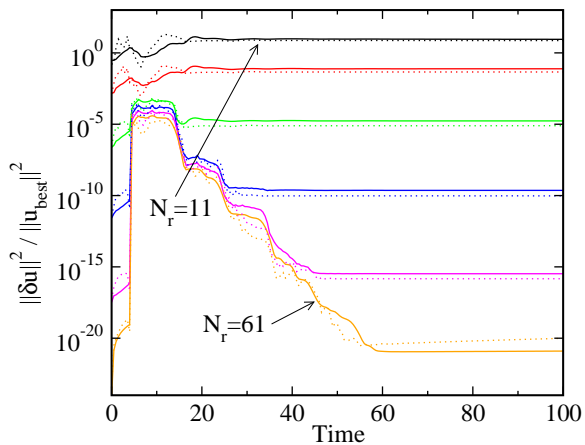}
\end{center}
\caption{Convergence test for fat Maxwell on $S^2\times R$, with the
signs of $\gamma_a$ switched at $t=4$ for all resolutions.  Shown are
norms of differences between solutions at different resolutions: solid
curves use $L^2$ norms and dotted curves use $L^\infty$ norms.  Compare to
runs with fixed $\gamma_a$ in
Fig.~\ref{f:SphereFreezingBcMinusConvergence}.}
\label{f:SphereControlForce4Convergence}
\end{figure}

\begin{figure}
\begin{center}
\includegraphics[width=3.0in]{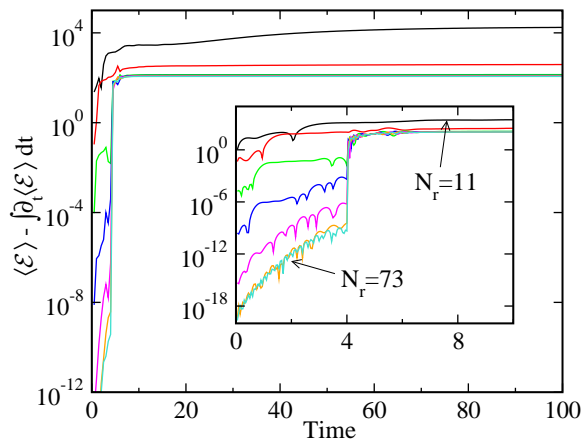}
\end{center}
\caption{Violation of constraint energy evolution equation,
Eq.~(\ref{e:denergynorm}), for cases with $\gamma_a$ switched at $t=4$
for all resolutions. Plotted is the difference between $\langle {\cal E}
\hs\rangle$ and the time integral of the right side of
Eq.~(\ref{e:denergynorm}) for each resolution.  Inset shows detail at
early times, showing that Eq.~(\ref{e:denergynorm}) is satisfied until
the switch at $t=4$.}
\label{f:SphereControlForce4BdryRelation}
\end{figure}

One effect that can destroy convergence in these tests is the fact
that the control mechanism is applied independently for each
resolution (as pointed out by Tiglio~\cite{Tiglio2003a}).  Therefore,
at a given value of $t$, the evolution equations used for one
resolution can be different (because of the sign of $\gamma_1$ and
$\gamma_2$) than the equations used for another resolution at that
time.  This effect should have significant consequences only on
quantities computed using different resolutions, such as the
differences plotted in Fig.~\ref{f:SphereControlConvergence}.  But
quantities computed using a single resolution, such as $\langle {\cal
E}\hs\rangle$, should not be affected.  When these latter quantities
are compared for different resolutions on the same plot, as in
Figs.~\ref{f:SphereControlConstraint}
and~\ref{f:SphereControlBdryRelation}, the graph will not look like
the ``classic'' convergence test in which all curves have the same
shape.  But the curves should (if everything else in the method is
convergent) decrease at roughly the correct rate.  Because we lose a
great deal of accuracy in Figs.~\ref{f:SphereControlConstraint}
and~\ref{f:SphereControlBdryRelation} compared to their uncontrolled
counterparts, Figs.~\ref{f:SphereFreezingBcConstraint}
and~\ref{f:SphereFreezingBcBdryRelation}, we believe that the fact
that the control mechanism is applied independently for each
resolution is not the primary cause of the problem.  In order to
eliminate this effect on convergence, we repeated our simulations, but
this time switched the sign of $\gamma_a$ only once at the time $t=4$
for each resolution, regardless of the sign of the right side of
Eq.~(\ref{e:denergynorm}) or the magnitude of $\langle {\cal
E}\hs\rangle$.  In this case, exactly the same evolution equations are
being solved at each resolution.  As shown in
Fig.~\ref{f:SphereControlForce4Convergence}, the convergence rate is
severely reduced even in this case when the signs of $\gamma_a$ are
switched at $t=4$.  Furthermore as shown in
Fig.~\ref{f:SphereControlForce4BdryRelation},
Eq.~(\ref{e:denergynorm}) is violated after the signs are switched.

\begin{figure}
\begin{center}
\includegraphics[width=3.0in]{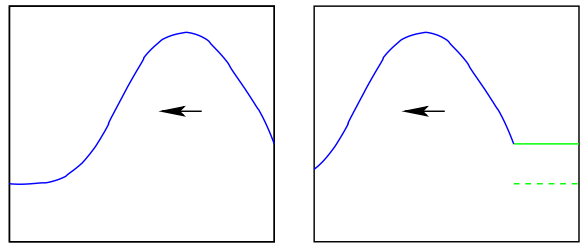}
\end{center}
\caption{Left curve represents a characteristic field at one instant
of time, and right curve the evolution of this field at a later time.
Freezing boundary conditions produce the non-smooth but continuous
solid curve extension, while standard maximally dissipative boundary
conditions produce the discontinuous dashed curve extension. }
\label{f:NonSmoothEvol}
\end{figure}

We now believe that this nonconvergence is caused by a lack of
smoothness of the fields that is introduced by the discontinuous
change in the evolution equations: Suppose the signs of $\gamma_a$ are
switched at a time $t_0$, and suppose that at time just before this
switch ($t=t_0-\epsilon$) some outgoing characteristic fields at the
boundary are nonzero.  When the signs of $\gamma_a$ are switched, some
of the outgoing and zero-speed characteristic fields will be converted
to incoming characteristic fields, and vice versa, as can be seen from
Eqs.~(\ref{e:Z1})--(\ref{e:U2}). (For example, switching the signs of
the $\gamma_a$ in Eq.~[\ref{e:U2}] while keeping $E_i$ and $D_{ij}$
fixed yields $U^{2\pm}_{\rm after} = -U^{2\mp}_{\rm before}$.)
Therefore, at a time just after this switch ($t=t_0+\epsilon$) the
solution will contain some non-vanishing incoming characteristic
fields near the boundary. However, our freezing boundary condition
requires the incoming characteristic fields to be constant in time at
the boundary.  Since the incoming characteristic fields propagate
inward, at times after the switch ($t>t_0+\epsilon$) a kink will
appear in the profile of these incoming fields.  This type of
boundary-condition-induced kink is illustrated in
Fig.~\ref{f:NonSmoothEvol}.  The sketch on the left in
Fig.~\ref{f:NonSmoothEvol} illustrates an incoming characteristic
field at time just after the switch ($t_0+\epsilon$), and the sketch
on the right shows the kink in this field that develops from the
boundary condition.  The existence of such a kink in the evolution
fields greatly reduces the convergence rate of our spectral evolution
method.  And even for finite-difference methods such a kink is likely
to reduce the convergence as well, since a kink in the fundamental
quantities translates into a discontinuity in the constraints.  Unless
great care is taken to ensure that discontinuous solutions are treated
properly (a standard problem in computational fluid dynamics but quite
foreign to vacuum numerical relativity because the gravitational field
is not expected to have physical shocks), Eq.~(\ref{e:denergynorm}) is
likely to be violated and the constraint preserving mechanism will
fail.  We have made several attempts to replace the freezing boundary
condition with a condition that smoothly adjusts the value of the
incoming fields at the boundary.  Unfortunately none of these attempts
have been very successful.

%
%
%

\subsection{Constraint-Preserving Boundary Conditions}
\label{s:bccontrol}

\begin{figure}
\begin{center}
\includegraphics[width=3.0in]{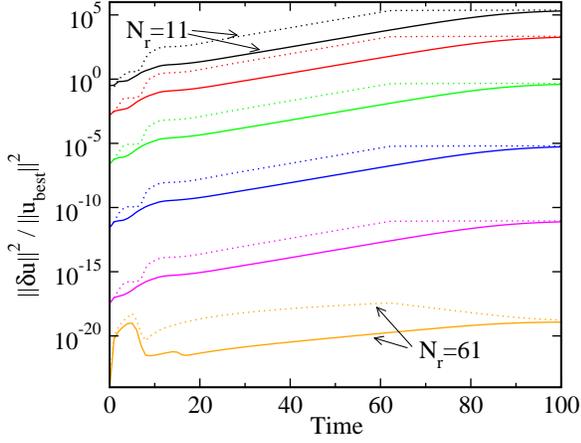}
\end{center}
\caption{Convergence test for fat Maxwell on $S^2\times R$, using
constraint-preserving boundary conditions and $\gamma_a<0$.
Shown are norms of differences between
solutions at different resolutions: solid curves use $L^2$ norms and
dotted curves use $L^\infty$ norms.}
\label{f:SphereConstraintBcConvergence}
\end{figure}

\begin{figure}
\begin{center}
\includegraphics[width=3.0in]{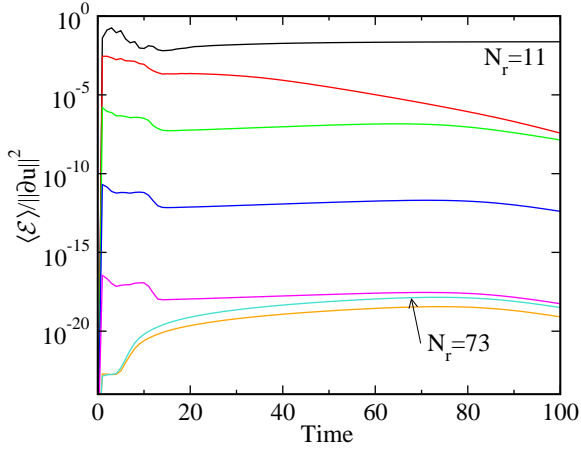}
\end{center}
\caption{Constraint violation for fat Maxwell on $S^2\times R$ using
constraint-preserving boundary conditions and $\gamma_a<0$.  Shown is
the constraint energy $\langle {\cal E} \hs\rangle$ divided by the norm
of the derivatives of the fundamental variables.}
\label{f:SphereConstraintBcConstraint}
\end{figure}

\begin{figure}
\begin{center}
\includegraphics[width=3.0in]{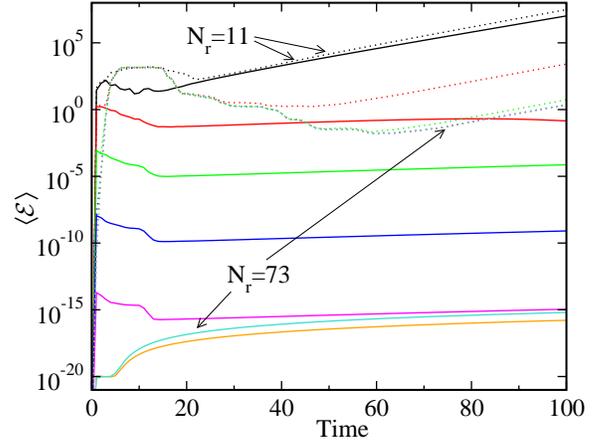}
\end{center}
\caption{Constraint violation for fat Maxwell on $S^2\times R$ for
$\gamma_a<0$. Solid curves use constraint-preserving boundary
conditions while dotted curves (same as the dotted curves in
Fig.~\ref{f:SphereFreezingBcConstraints}) use freezing boundary
conditions.  Seven different resolutions are depicted for each type of
boundary condition, but only the lowest resolution curves are distinct
at the scales shown.}
\label{f:SphereConstraintsCompareBc}
\end{figure}

Finally, we have performed a series of tests on the constraint
preserving boundary conditions described in
Sec.~\ref{s:boundaryconditions}.
Figures~\ref{f:SphereConstraintBcConvergence}
and~\ref{f:SphereConstraintBcConstraint} show evolutions of our
dynamical initial data on $S^2\times R$ (analogous to that used in
Figs.~\ref{f:SphereFreezingBcMinusConvergence}
and~\ref{f:SphereFreezingBcMinusConstraint}) with the boundary
condition on $U^{2-}$ now set according to the constraint preserving
condition Eq.~(\ref{eq:ConstraintBc}).  The $\gamma_a$ are negative
for the plots in Figs.~\ref{f:SphereConstraintBcConvergence}
and~\ref{f:SphereConstraintBcConstraint}.  The constraints are
satisfied, and the simulation appears to be convergent (except for a
late-time angular numerical instability, not visible on the plots,
that appears identical to the numerical instability discussed at the
end of Section~\ref{s:nocontrol}).  We have also performed these
evolutions using positive values of $\gamma_a$, and the simulations
appear to be convergent, completely stable, and constraint-preserving
in this case.  Figure~\ref{f:SphereConstraintsCompareBc} compares the
unnormalized constraint energy for these evolutions with those run
with freezing boundary conditions.  Thus adopting constraint
preserving boundary conditions clearly does improve the constraint
preserving properties of these evolutions much more than the active
constraint control method.

\begin{figure}
\begin{center}
\includegraphics[width=3.0in]{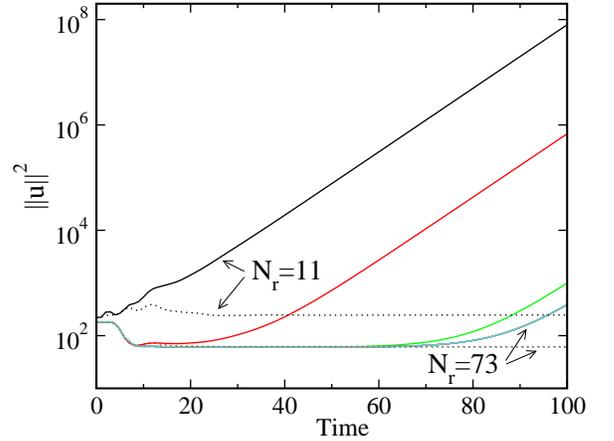}
\end{center}
\caption{Norm of the fundamental variables
for fat Maxwell on $S^2\times R$ with
$\gamma_a<0$ (solid) and $\gamma_a>0$ (dotted), using constraint-preserving
boundary conditions. Even though the constraints are satisfied
for $\gamma_a<0$, the fundamental quantities increase exponentially, but
in a convergent manner. Seven different resolutions are depicted for each case,
but only the lowest resolution curves are distinct at the scales shown.}
\label{f:SphereBlowup}
\end{figure}

But this is not the entire story.
Figure~\ref{f:SphereBlowup} shows the norm of the fundamental
dynamical fields $||u||^2$ for evolutions using constraint preserving
boundary conditions with $\gamma_1=-0.1$ (solid curves) and
$\gamma_1=0.1$ (dotted curves).  This plot shows that while the
positive $\gamma_1$ evolutions are stable, those with negative
$\gamma_1$ are not.  A more extensive sampling of the parameter space
reveals that evolutions preformed with
$\gamma_1=1/\gamma_2=\{0.1,1.0,2.5,2.9\}$ (for which the principal
evolution system is symmetric hyperbolic) appear to all be convergent,
constraint preserving, and stable.  Conversely, we find that
evolutions performed with $\gamma_1=1/\gamma_2=\{-1.0,-0.1,3.5,4.1\}$
(for which the principal evolution system is strongly but not
symmetric hyperbolic) are all constraint preserving but unstable.
These evolutions are numerically convergent for the resolutions
and time scales we have tested (except for a slowly-growing
angular numerical
instability, not visible on Figure~\ref{f:SphereBlowup},
that appears at late times or for high
angular resolutions).  Therefore, the type of growth
seen in Figure~\ref{f:SphereBlowup} appears to represent a solution of
the partial differential equations.  Since these solutions do satisfy
the constraints, the driving force for these instabilities must be an
excess of incoming radiation that is reflected back into the
computational domain by the boundary condition.  We refer to this
type of instability as a {\it boundary condition driven
instability}. Thus the constraint preserving boundary conditions are a
great improvement over the other methods studied here, but they do not
completely eliminate all the instabilities in these strongly
hyperbolic cases. Further study will be needed to determine whether
these boundary conditions can be improved.

%
%
%
%

\section{Discussion}
\label{s:discussion}

This paper explores the effectiveness of two methods for controlling
the growth of constraints in hyperbolic evolution systems.  Using an
expanded version of the Maxwell evolution system---which we call the
fat Maxwell system---we showed that significant constraint violations
and in some cases even constraint violating instabilities occur when
the evolutions are performed using ``standard'' numerical methods and
boundary conditions.  We show that the active constraint control
method (which has been studied by Tiglio and his
collaborators~\cite{Tiglio2003a,Tiglio2003b}) is not very effective in
controlling the growth of constraints in the fat Maxwell system when
spectral numerical methods are used.  This lack of effectiveness
appears to be caused by the non-smooth nature of the control mechanism
for this system.  We also show that constraint preserving boundary
conditions are very effective in suppressing the constraint violations
in this system.  Unfortunately these constraint preserving boundary
conditions did not eliminate the instabilities for the strongly (but
not symmetric) hyperbolic evolution systems.  In these cases these
boundary conditions merely converted a constraint violating
instability into a boundary condition driven instability.
Generalizing these methods to more complicated systems like the
Einstein evolution equations should be straightforward.  For more
general systems the analogue of the constraint energy evolution
equation, Eq.~(\ref{eq:ConstraintBc}), will contain both boundary
terms like the fat Maxwell system and also volume terms, so that
constraint violations can be generated both at boundaries and in the
bulk.  We expect that constraint preserving boundary conditions alone
will not be sufficient to control the constraint violating
instabilities that occur in these more general systems.  Instead we
expect that some combination of methods will be needed.  The
disappointing results obtained here with the active constraint control
mechanism suggest that significant improvements will be needed to make
this method useful for helping to control the growth of constraints in
the Einstein system for evolutions based on spectral methods.

\acknowledgments We thank Michael Holst, Oscar Reula,
Olivier Sarbach and Manuel Tiglio for helpful
discussions concerning this work.  This work was supported in part by
NSF grants PHY-0099568, PHY-0244906 and NASA grants NAG5-10707,
NAG5-12834 at Caltech, and NSF grants PHY-9900672, PHY-0312072 at
Cornell.

\bibstyle{prd}
\bibliography{References}

\end{document}